\documentclass[twocolumn,english,aps,superscriptaddress,floatfix,final,showpacs,prb,eqsecnum]{revtex4}
\usepackage{amsmath}
\usepackage{graphicx}
\usepackage{amssymb}

\makeatletter
\usepackage{epsfig}
\usepackage{dcolumn}
\usepackage{bm}
\usepackage{amsfonts}
\usepackage{graphicx}
\begin{document}

\title{Effective model of the electronic Griffiths phase}

\author{D. Tanaskovi\'{c}}

\affiliation{Department of Physics and National High Magnetic Field Laboratory, Florida State
University, Tallahassee, Florida 32306, USA.}

\author{E. Miranda}

\affiliation{Instituto de Fisica Gleb Wataghin, Unicamp, Caixa Postal 6165, 
Campinas, SP, CEP 13083-970, Brazil.}

\author{V. Dobrosavljevi\'{c}}

\affiliation{Department of Physics and National High Magnetic Field Laboratory, Florida State
University, Tallahassee, Florida 32306, USA.}

\date{\today{}}

\begin{abstract}
We present simple analytical arguments explaining the universal
emergence of electronic Griffiths phases as precursors of
disorder-driven metal-insulator transitions in correlated
electronic systems. A simple effective model is constructed and
solved within Dynamical Mean Field Theory. It is shown to
capture all the qualitative and even quantitative aspects of such
Griffiths phases.
\end{abstract}


\maketitle

\section{Introduction}

The recent discovery of a number of heavy fermion materials with
non-Fermi liquid (NFL) thermodynamic and transport properties has
been followed by a significant theoretical effort to understand
the origin of the NFL behavior.\cite{stewartNFL} In the cleaner
systems, the proximity to a quantum critical point appears to
dominate many of the observed
properties.\cite{lohneysen,groscheetal,hertz,moryia,japiassuetal,millis,pierspepinsirevaz,schroedernature,sietal}
In other heavy fermion systems disorder seems to play a more
essential role and appears to be crucial for understanding the NFL
behavior.\cite{bernaletal,boothetal,buttgenetal,dougetal,dougetal2,baueretal,boothetal2}
Many experiments can be explained by the disordered Kondo
model,\cite{bernaletal} which has recently been put on a much
stronger microscopical
foundation.\cite{mirandavlad1,mirandavlad2,aguiaretal1}

The emergence of electronic Griffiths phases in models of correlated
electrons has been
established\cite{vladgabisdmft1,mirandavlad1,mirandavlad2,aguiaretal1}
as a universal phenomenon, within a class of extended
(``statistical'') Dynamical Mean Field Theory (DMFT)
approaches.\cite{vladgabisdmft1} This \textit{stat}DMFT method
provides an exact (numerical) treatment of localization in the absence
of interactions, and reduces to the standard DMFT
equations\cite{georgesrmp} in the absence of disorder. When both
interactions and localization are present, non-Fermi liquid behavior
emerges universally,\cite{mirandavlad1}
as a precusor of a disorder-driven metal-insulator transition,
due to a very broad
distribution $P(T_{K})$ of local Kondo temperatures. This distribution
has a low-$T_{K}$ tail of the form $P(T_{K})$ $\sim
T_{K}^{\,\,\,\alpha-1}$, independent of the microscopic details or the
specific form of disorder. The exponent $\alpha=\alpha(W)$ is found to
be a smooth, monotonically decreasing function of the disorder
strength $W$, and the NFL behavior emerges for
$W$ greater than the
critical value $W_{nfl}$ corresponding to $\alpha\leq1$, when
$P(T_{K})$ becomes singular at small $T_{K}$.  As in other Griffiths
phases, the thermodynamic and transport properties in this NFL region
are dominated by rare events, which in this model correspond to sites
with the lowest Kondo temperatures.

In this paper, we show that the same behavior is found in a
simpler, standard DMFT version of the model with a judicious
choice of bare disorder. We should emphasize that localization is
not present in this effective model, but the Griffiths phase
emerges in qualitatively the same fashion as in the above more
realistic calculations. We discuss how the specific disorder
distribution which is hand-picked in the effective model is
dynamically generated by fluctuation effects within the
\textit{stat}DMFT formulation, elucidating the origin of the
universality of the Griffiths phase behavior. In addition, the
simplicity of this DMFT effective model makes it possible to
describe all the qualitative features of the solution using simple
analytical arguments, thus eliminating the need for large scale
numerical computations in the description of the electronic
Griffiths phase. This may be crucial in order to address more
complicated issues, such as the role of additional
Ruderman-Kittel-Kasuya-Yosida (RKKY) interactions in disordered
Kondo alloys.\cite{dougetalreview}

This paper is organized as follows. Sec.~\ref{sec:Model} introduces the effective
model for the electronic Griffiths phase as a DMFT model with a Gaussian distribution
of random site energies. This model is solved analytically in the Kondo limit in
Sec.~\ref{sec:Analytical-solution}, and numerically in Sec.~\ref{sec:Numerical-results}.
The arguments explaining the universal aspects of the form of the renormalized disorder
are presented in Sec.~\ref{sec:Role-of-fluctuations}. Sec.~\ref{sec:Electronic-Griffiths-phase}
establishes a connection with the Griffiths phase in a single band Hubbard model,
and Sec.~\ref{sec:Summary-and-outlook} contains our conclusions.

\section{Model}

\label{sec:Model}

We consider the Anderson lattice model where the disorder is
introduced by random site energies $\varepsilon_{i}$ in the
conduction band, as given by the Hamiltonian
\begin{eqnarray}
H& = &-t\sum_{\langle ij \rangle \sigma}(c_{i\sigma}^{\dagger}c_{j\sigma}
+\mbox{H.c.}) +
 \sum_{j\sigma}(\varepsilon_j- \mu )c_{j\sigma}^{\dagger}c_{j\sigma}
\nonumber \\
&&+
V\sum_{j\sigma}\left(c_{j\sigma}^{\dagger}f_{j\sigma}
+
\mathrm{H}.\mathrm{c}.\right)
+ 
\sum_{j\sigma}E_{f}f_{j\sigma}^{\dagger}f_{j\sigma}
\nonumber \\
&&+U\sum_{j}f_{j\uparrow}^{\dagger}f_{j\uparrow}
f_{j\downarrow}^{\dagger}f_{j\downarrow},\label{2.1}
\end{eqnarray}
%
%
where $f_{j\sigma}$ and $c_{j\sigma}$ are annihilation operators for
$f$- and conduction electrons, respectively. $V$ is the hybridization
parameter, and $E_{f}$ is the $f$-electron energy. We assume
$U\rightarrow\infty,$ and choose a Gaussian distribution of random
site energies for the conduction band \begin{equation}
P(\varepsilon_{i})=(2\pi
W^{2})^{-1/2}\exp\{-\frac{1}{2}\varepsilon_{i}^{2}/W^{2}\}.\label{2.2}\end{equation}
In Sec.~\ref{sec:Role-of-fluctuations} we will explain how this
particular disorder distribution comes out naturally from the more
generic \textit{stat}DMFT approach.

To solve these equations, we use the DMFT
approach,\cite{georgesrmp} which is formally exact in the limit of
large coordination. We concentrate on a generic unit cell $j$,
containing an $f$-site and its adjoining conduction electron
Wannier state. After integrating out the conduction electron
degrees of freedom, we obtain the effective action for the
$f$-electron on site $j$\begin{eqnarray}
S_{imp}(j) & = & \sum_{\sigma}\int_{0}^{\beta}d\tau\int_{0}^{\beta}d\tau'f_{j\sigma}^{\dagger}(\tau)\left[\delta(\tau-\tau')(\partial_{\tau}+E_{f})\right.\nonumber \\
 & + & \left.\Delta_{fj}(\tau-\tau')\right]f_{j\sigma}(\tau').\label{2.3}\end{eqnarray}
 Here, the restriction of no double $f$-site occupancy is implied. The
hybridization function $\Delta_{fj}$ between the $f$-electron and
the conduction bath $\Delta_{c}$ is given by \begin{equation}
\Delta_{fj}(i\omega_{n})=\frac{V^{2}}{i\omega_{n}-\varepsilon_{j}+\mu-\Delta_{c}(i\omega_{n})}.\label{2.4}\end{equation}
 The self-consistency condition for the conduction bath (cavity field) assumes a
simpler form for the semi-circular model density of
states\cite{georgesrmp}, which we use for simplicity. All the
qualitative features of our solution are independent of the the
form of the lattice, and the quantitative results depend only
weakly on the details of the electronic band structure. For this
model
$\Delta_{c}(i\omega_{n})=t^{2}\overline{G_{c}}(i\omega_{n})$,
where $\overline{G_{c}}(i\omega_{n})$ is the disorder-averaged Green's
function of the conduction electrons, and the self-consistency is
enforced by \begin{equation}
\overline{G_{c}}(i\omega_{n})=\langle[i\omega_{n}-\varepsilon_{j}+\mu-t^{2}\overline{G_{c}}(i\omega_{n})-\Phi_{j}(i\omega_{n})]^{-1}\rangle,\label{2.5}\end{equation}
 where \begin{equation}
\Phi_{j}(i\omega_{n})=\frac{V^{2}}{i\omega_{n}-E_{f}-\Sigma_{fj}(i\omega_{n})},\label{2.6}\end{equation}
 and $\Sigma_{fj}$ is the $f$-electron self-energy derived from the impurity action
of Eq.~(\ref{2.3}). {}From a technical point of view, within DMFT
the solution of the disordered Anderson lattice problem reduces to
solving an ensemble of a single impurity problems supplemented by
a self-consistency condition.

We will solve the system of Eqs.~(\ref{2.3})-(\ref{2.6}) at zero
temperature using the slave boson mean field theory
approach.\cite{readnewns2,colemanlong} This approximation is
known\cite{mirandavlad1,mirandavlad2,aguiaretal1} to reproduce all
the qualitative and even most of the accurately quantitative
features of the exact DMFT solution at $T=0$. It introduces
renormalization factors (quasi-particle weights) $Z_{j}$ and
renormalized $f$-energy levels $\varepsilon_{fj}$, which are
site-dependent quantities in the case of a disordered lattice.
These parameters are determined by the saddle-point slave boson
equations (see Ref.~\onlinecite{mirandavladgabi1} for more
details) which, on the real frequency axis, assume the form
\begin{equation}
-\frac{1}{\pi}\int_{-\infty}^{0}d\omega\,\mathrm{Im}\left[\frac{1}{\omega-\varepsilon_{fj}-Z_{j}\Delta_{fj}(\omega)}\right]=\frac{1}{2}(1-Z_{j}),\label{2.7}\end{equation}
\begin{equation}
\frac{1}{\pi}\int_{-\infty}^{0}d\omega\,\mathrm{Im}\left[\frac{\Delta_{fj}(\omega)}{\omega-\varepsilon_{fj}-Z_{j}\Delta_{fj}(\omega)}\right]=\frac{1}{2}(\varepsilon_{fj}-E_{f}).\label{2.8}\end{equation}
 Eq.~(\ref{2.6}) in this case becomes \begin{equation}
\Phi_{j}(\omega)=\frac{Z_{j}V^{2}}{\omega-\varepsilon_{fj}}.\label{2.9}\end{equation}

\section{Analytical solution in the Kondo limit}

\label{sec:Analytical-solution}

Before presenting a numerical solution of the slave boson Eqs.~(\ref{2.7})-(\ref{2.8})
supplemented by the self-consistency condition of Eq.~(\ref{2.5}), we will solve
these equations analytically in the Kondo limit for a \emph{given} conduction bath.
A comparison with the numerical solution will show that the self-consistency does
not qualitatively change the analytical results.

The slave boson equations simplify in the Kondo limit $Z_{j}\rightarrow0$. The integral
in Eq.~(\ref{2.7}) is dominated by the low-frequency region, and the frequency
dependence in $\Delta_{c}$ and $\Delta_{fj}$ can be neglected. Therefore, after
integration \begin{equation}
\varepsilon_{fj}\approx -Z_{j}\mathrm{Re}\left[\Delta_{fj}(0)\right],\label{3.1}\end{equation}
 where, for simplicity, we took a semi-circle conduction bath with $\mu=0$. In
the integral of Eq.~(\ref{2.8}), the frequency dependence of
$\Delta_{fj}$ can also be neglected. Introducing the energy
cutoff $D$ and using Eq.~(\ref{3.1}) we obtain
\begin{eqnarray}
Z_{j} & \approx & D\frac{\varepsilon_{j}^{2}+(\pi t^{2}\rho_{o})^{2}}{\pi t^{2}\rho_{o}V^{2}}e^{-\pi^{2}t^{2}\rho_{o}/J}e^{-\varepsilon_{j}^{2}/t^{2}\rho_{o}J}e^{\varepsilon_{j}/2t^{2}\rho_{o}}\nonumber \\
 & = & Z_{o}\frac{\varepsilon_{j}^{2}+(\pi t^{2}\rho_{o})^{2}}{(\pi t^{2}\rho_{o})^{2}}\exp\left[-\frac{\varepsilon_{j}^{2}}{t^{2}\rho_{o}J}\left(1-\frac{J}{2\varepsilon_{j}}\right)\right].\label{3.2}\end{eqnarray}
 Here, $\rho_{o}$ is the density of states (DOS) of the conduction electrons at
the Fermi level, $J=2V^{2}/|E_{f}|$, and $Z_{o}=Z\left(\varepsilon_{j}=0\right)$.
The Kondo temperature is proportional to the quasi-particle weight, $T_{Kj}=\pi V^{2}\rho_{o}Z_{j}$.
In the limit $\varepsilon_{j}\gg J/2$ and neglecting a weak site-energy dependence
in the pre-factor, we obtain\begin{equation}
T_{Kj}\approx T_{K}^{0}e^{-1/\lambda_{j}},\label{3.3}\end{equation}
 where the site dependent coupling constant is\begin{equation}
\lambda_{j}=\frac{t^{2}\rho_{o}J}{\varepsilon_{j}^{2}},\label{3.4}\end{equation}
 and $T_{K}^{0}$ is the Kondo temperature in the clean limit (for $\varepsilon_{j}=0$).
{}From these equations, we can immediately find the desired distribution of local Kondo
temperatures $P(T_{K})=P(\varepsilon(T_{K}))|d\varepsilon/dT_{K}|$, which (up to
a negligible logarithmic correction) is given asymptotically by \begin{equation}
P(T_{K})\propto(T_{K}/T_{K}^{0})^{\alpha-1},\label{3.5}\end{equation}
 with \begin{equation}
\alpha(W)=\frac{t^{2}\rho_{o}J}{2W^{2}}.\label{3.6}\end{equation}
 This expression is one of the central results of this paper. It has exactly the form expected for a Griffiths
 phase, where the exponent characterizing the local energy scale
 distribution assumes a parameter-dependent (tunable) form.

To show how the NFL behavior appears due to the singularity in $P(T_{K})$, we use
the standard expression due to Wilson for the magnetic susceptibility\cite{hewson}\begin{equation}
\chi(T,T_{K})=\frac{C}{T+aT_{K}},\label{3.7}\end{equation}
 which is an excellent approximation for a single Kondo impurity. Here $C$ and $a$
are constants. In the disordered case, we can split the average susceptibility $\chi(T)=\int_{0}^{\infty}P(T_{K})\chi(T,T_{K})dT_{K}$
in a regular {}``bulk'' part\begin{equation}
\chi_{r}(T)=\int_{\Lambda}^{\infty}P(T_{K})\chi(T,T_{K})dT_{K},\label{3.8}\end{equation}
 and a potentially singular part \begin{equation}
\chi_{s}(T)=C_{1}\int_{0}^{\Lambda}\frac{T_{K}^{\alpha-1}}{T+aT_{K}}dT_{K},\label{3.9}\end{equation}
 coming from the tail with low Kondo temperatures ($\Lambda$ is a crossover scale).
At weak disorder, the exponent $\alpha$ is large and the distribution $P(T_{K})$
is regular, $\chi(0)=\chi_{o}+C_{2}/(\alpha-1)$, but NFL behavior emerges once $\alpha\leqslant1$,
which corresponds to \begin{equation}
W\geq W_{nfl}=\sqrt{t^{2}\rho_{o}J/2}.\label{3.10}\end{equation}
 For $\alpha=1$ the magnetic susceptibility has a logarithmic divergence, $\chi(T)\propto\ln(1/T)$,
characteristic of marginal Fermi liquid behavior,\cite{varmamfl} while for $\alpha<1$
a power law divergence is obtained, $\chi(T)\propto T^{\alpha-1}$ as $T\rightarrow0$.
The same singularity also leads to an anomalous behavior in the transport properties,
as shown in detail in Refs.~\onlinecite{mirandavladgabi1} and \onlinecite{mirandavladgabi2}.

\section{Numerical results}

\label{sec:Numerical-results}

\begin{figure}
\begin{center}\includegraphics[%
  width=2.919in,
  keepaspectratio]{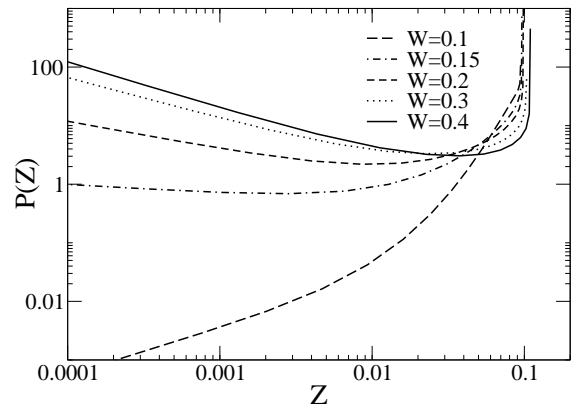}\end{center}

\caption{\label{fig1} Distribution of the local Kondo temperatures. The disorder ranges
from moderate $W=0.1$ to strong $W=0.4$. We used $V=0.5$, $E_{f}=-1$, and $\mu=-0.1$.}
\end{figure}

In the above derivation we ignored the fact that the conduction bath $\Delta_{c}$
has to be self-consistently determined. This will also produce particle-hole asymmetry
and an asymmetric distribution of Kondo temperatures $T_{Kj}$. A nonzero chemical
potential will further increase this asymmetry. However, the numerical solution we
obtained using the slave boson approximation at zero temperature shows that the essential
physics described by Eqs.~(\ref{3.3})-(\ref{3.6}) remains qualitatively correct.
The distribution of local Kondo temperatures in the asymptotic limit is indeed a
power law, $P(T_{K})\sim T_{K}\,\!^{\alpha-1}$, where the exponent $\alpha$ is
a decreasing function of disorder.

Fig.~\ref{fig1} shows the distribution $P(T_{K})$ for several values
of the disorder distribution strength $W$. For the parameters that we
here use, the system is close to the Kondo limit, and the Kondo gap of
the clean system is approximately 0.04 (in energy units of the half
bandwidth of bare DOS). The NFL behavior appears for $W\gtrsim0.14$.
We note that in the NFL regime the power law behavior appears already
for the site energies $\varepsilon_{j}$ which deviate only moderately
from the mean (zero) value.  In other words, the asymptotic behavior
is established \emph{well before} we attain very rare realizations of
$\varepsilon_{j}$ which belong to the tail of the Gaussian
distribution. For example, for $W=0.3$, sites with
$\varepsilon_{j}\gtrsim 0.4$ (which correspond to $Z\lesssim 0.01$)
are already in the power-law regime.

\begin{figure}[h]
\begin{center}\includegraphics[%
  width=2.919in,
  keepaspectratio]{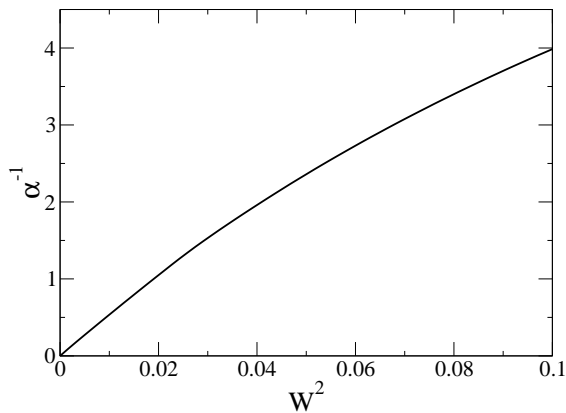}\end{center}
\caption{\label{fig2}Inverse power law parameter $\alpha^{-1}$ as a function of $W^{2}$. For
weak and moderate disorder this dependence is linear. Here $V=0.5$, $E_{f}=-1$,
and $\mu=-0.2$. }
\end{figure}

\begin{figure}
\begin{center}\includegraphics[%
  width=2.919in]{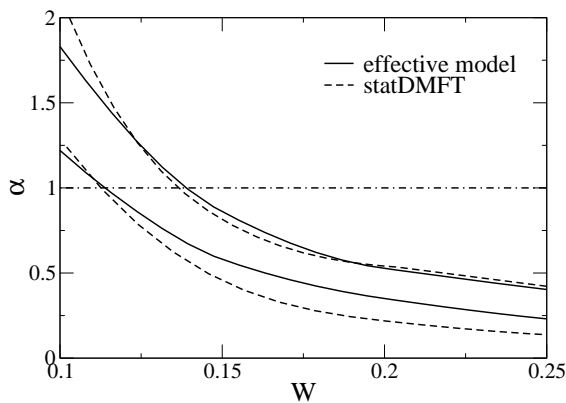}\end{center}

\caption{\label{fig3}Power law exponent $\alpha$ as a function of disorder strength measured by
the standard deviation $W$. Full lines are the effective model results, and dashed
lines are the \emph{stat}DMFT results. The hybridization $V$ is taken to be 0.5
and $E_{f}=-1$. The chemical potential is $\mu=-0.1$ (upper lines), and $-0.5$
(lower lines). The onset of NFL behavior occurs at $\alpha=1$. }
\end{figure}

According to the simplified derivation from
Sec~\ref{sec:Analytical-solution}, the exponent $\alpha$ is
inversely proportional to $W^{2}$. The numerical results shown in
Fig.~\ref{fig2} confirm such behavior for weak and moderate
disorder. For strong disorder there appear some deviations from
this formula, which can be ascribed to the dependence of the DOS at
the Fermi level on the disorder strength.

Before we present arguments which justify our effective DMFT model
approach, let us make a direct comparison with the \emph{stat}DMFT
results from Ref.~\onlinecite{aguiaretal1}. In this approach, very
broad distributions of local Kondo temperatures are generated for
\textit{arbitrary} distributions of bare disorder.
In particular, even if the bare distribution is bounded, sites
with arbitrarily small Kondo temperatures will exist, and their
distribution will have a power law tail. This is a consequence of
the spatial fluctuations of the conduction electron cavity field,
as we discuss in detail in the next Section. In Fig.~\ref{fig3} we
compare the values of the exponent $\alpha$ for the effective
model with Gaussian disorder of variance $W^2$, and the
\emph{stat}DMFT results obtained for a \emph{bounded} uniform
distribution of bare disorder with the same variance. Remarkably,
not only does the electronic Griffiths phase emerge in the same
fashion, but the numerical values of disorder strength determining
the onset of NFL behavior are also almost the same. The comparison
is made for two different values of the chemical potential. As we move
further away from half-filling by changing the chemical potential,
the critical value $W_{nfl}$ decreases. That is expected since
$\alpha$ should be proportional to the bare (noninteracting) DOS
at the Fermi level.

\section{The role of spatial fluctuations and the form of renormalized disorder}

\label{sec:Role-of-fluctuations}

In this Section we explain the universal aspects of the emergence of the electronic
Griffiths phase within the more generic statistical DMFT. In particular, we show
how the Gaussian tails in the distribution of renormalized disorder appear for an
arbitrary form of the bare disorder. Moreover, we present arguments showing that
the Griffiths phase appears generically as a precursor of the Mott-Anderson metal-insulator
transition.

\subsection{Universality of the renormalized disorder distribution}

In the above DMFT formulation, we had to choose a special form of disorder distribution
in order to obtain the desired power-law distribution of Kondo temperatures. Had
we chosen a different distribution, the results would not have held. For example,
for a bounded distribution of site energies, there would always be a minimum value
of the Kondo temperature, and thus no power-law tail. On the other hand, from numerical
simulations of lattices with finite coordination, it has been established that the
emergence of the Griffiths phase is a universal phenomenon.\cite{aguiaretal1} Why?
To understand the reason for this, we note that for finite coordination (as opposed
to the DMFT limit) the cavity bath $\Delta_{c}$ is not self-averaging, but is a
site-dependent, random quantity $\Delta_{cj}$. In this \emph{stat}DMFT formulation,
the local conduction electron Green's function is given by \begin{equation}
G_{cj}(i\omega_{n})=\frac{1}{i\omega_{n}-\varepsilon_{j}+\mu-\Delta_{cj}(i\omega_{n})-\Phi_{j}(i\omega_{n})},\label{4.1}\end{equation}
 where $\Phi_{j}$ describes the local scattering of the conduction electrons off
the $f$-shell at site $j$, and is given as before by Eq.~(\ref{2.6}).

For weak disorder, the corresponding fluctuations are small, and we can separate
\begin{equation}
\Delta_{cj}=\Delta_{c}^{av}+\delta\Delta_{cj}.\label{4.2}\end{equation}
 In the following, we compute the distribution function for the fluctuations of the
cavity field, which will lead to the renormalized form of the disorder distribution
function. The renormalized site energies can be defined as \begin{equation}
\widetilde{\varepsilon}_{j}=\varepsilon_{j}+\delta\Delta_{cj}^{R},\label{4.3}\end{equation}
 where $\delta\Delta_{cj}^{R}=\mathrm{Re}\left[\delta\Delta_{cj}(\omega=0)\right]$.
We stress that the cavity fluctuations are present for a general finite coordination
electronic system in the presence of disorder of any kind. In particular, the disorder
in hybridization parameters $V_{j}$, or local $f$-energy levels $E_{fj}$, will
induce fluctuations in the local DOS even if random site energies $\varepsilon_{j}$
in the conduction band are absent. Furthermore, as we argue in the next subsection,
the renormalized distribution $P(\widetilde{\varepsilon}_{j})$ will have universal
Gaussian tails even if the bare distribution $P(\varepsilon_{j})$ is bounded. Note
that $\delta\Delta_{cj}$ has a real as well as an imaginary part $\delta\Delta_{cj}^{I}$,
due to the fact that fluctuations locally violate particle-hole symmetry.
However, we show in the Appendix that $\delta\Delta_{cj}^{I}$ fluctuations, at least when
treated to leading order, do not produce singular behavior in $P(T_{K})$ and therefore
can be neglected when examining the emergence of the electronic Griffiths phase.

\subsection{The Gaussian nature of the renormalized distribution}

{}From detailed numerical studies it has become clear that the onset of the Griffiths
phase in disordered Anderson lattices generally occurs already for a relatively moderate
amount of disorder. In this limit, the relevant distributions are determined essentially
by the central limit theorem, therefore resulting in a Gaussian form of the tails
for $P(\widetilde{\varepsilon}_{j})$. This is precisely what is needed to justify
the DMFT effective model, where such Gaussian tails are assumed from the outset.

Before engaging in more precise computations of these
distributions, it is worth pausing to comment on the physical
validity of the assumed Gaussian statistics, i.e. the relevance of
the central limit theorem in the cases of interest. Quite
generally, if a certain quantity can be represented as a sum of a
large number of independent random variables, then the central
limit theorem tells us that the resulting distribution will be
Gaussian, irrespective of the specific form of the distributions
of the individual terms in the sum. In our case, the fluctuations
of the local cavity field result from Friedel oscillations of the
electronic wave functions, induced by other impurities which may
lie at a relatively long distance from the given site. This is a
result of the slow ($\sim R^{-d}$) decay of the amplitude of the
Friedel oscillations in $d$ dimensions, where $R$ is the distance
from the impurity site. The situation is very reminiscent of the
Weiss molecular field of an itinerant magnet, where the RKKY
spin-spin interactions have a long range character for the very
same reason, being as they are a reflection of similar Friedel
oscillations. Furthermore, as we will explicitly show, the leading
corrections (to order $\mathcal{O}\left(W^{2}\right)$) at weak
disorder take the form of a linear superposition of contributions
from single impurity scatterers, and thus of a sum of independent
random numbers, for which we expect the central limit theorem to
hold.

To obtain the precise form of this distribution, it therefore suffices to compute
the variance \begin{equation}
\sigma_{R}^{2}=\left\langle (\delta\Delta_{cj}^{R})^{2}\right\rangle ,\label{4.4}\end{equation}
 to leading order in disorder strength. To compute the fluctuations $\delta\Delta_{cj}$
at weak disorder, we note that the cavity field $\Delta_{cj}$ can be computed if
we consider a particular site with $\varepsilon_{j}=0$ (call it site $0$), and
compute its Green's function in a random medium. At zero frequency for this site
\begin{equation}
\Delta_{co}=\mu-1/G_{oo}.\label{4.5}\end{equation}
 The corresponding variation is \begin{equation}
\delta\Delta_{co}=\delta G_{oo}/(G_{oo})^{2}.\label{4.6}\end{equation}
 We still need to compute the fluctuation $\delta G_{oo},$ which can be expanded
in powers of the random potential $\varepsilon_{j}$. In doing this, we have ignored
the interaction renormalizations of the random potential for conduction electrons.
We will return to re-examine this effect in the Appendix. Note, however, that in
the absence of interactions in the environment of a given site, the following expressions
provide the exact leading contributions at weak disorder.

To leading order, we can write \begin{equation}
\delta G_{oo}=\sum\limits _{j}\varepsilon_{j}(G_{oj})^{2}+\mathcal{O}(\varepsilon^{2}).\label{4.7}\end{equation}
 This gives \begin{equation}
\sigma_{R}^{2}=CW^{2}+\mathcal{O}(W^{4}),\label{4.8}\end{equation}
 where \begin{equation}
C=\sum\limits _{\genfrac{}{}{0pt}{2}{j}{(j\neq0)}}\left[\mathrm{Re}\frac{(G_{oj})^{2}}{(G_{oo})^{2}}\right]^{2}.\label{4.9}\end{equation}
 The Green's function sum $C$ will numerically depend on the lattice geometry, but
will generally be a dimensionless number of order one.

\begin{figure}
\begin{center}\includegraphics[%
  width=2.919in,
  keepaspectratio]{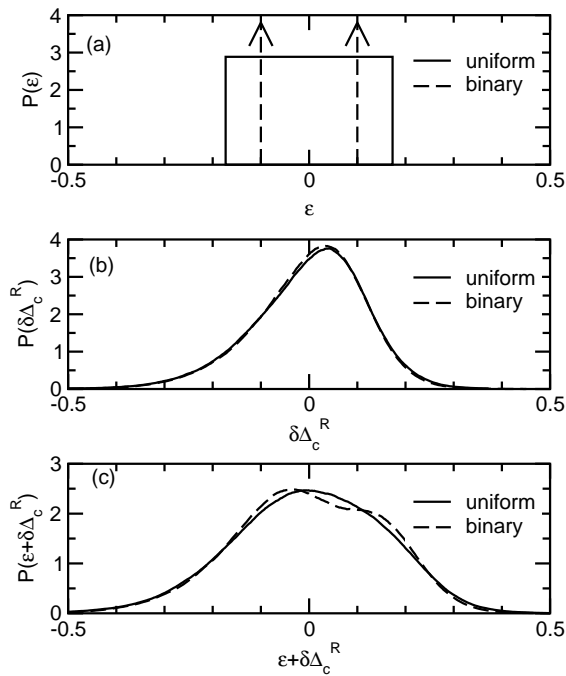}\end{center}

\caption{\label{fig4}\emph{Stat}DMFT results for the disorder distributions. Universal Gaussian-like
tails appear in the cavity field fluctuations, panel (b), and the renormalized disorder
distributions, panel (c), even though the bare disorder distributions are bounded
(uniform and binary), panel (a). We used $V=0.5$, $E_{f}=-1$, and $\mu=-0.5$. }
\end{figure}

The distribution of renormalized disorder $\widetilde{\varepsilon}=\varepsilon+\delta\Delta_{c}^{R}$
is given by a convolution of the distributions $P_{1}(\varepsilon)$ and $P_{2}(\delta\Delta_{c}^{R})$\begin{equation}
P(\widetilde{\varepsilon})=\int_{-\infty}^{\infty}d\omega\, P_{1}(\widetilde{\varepsilon}-\omega)P_{2}(\omega).\label{4.10}\end{equation}
 If the bare distribution is bounded, (e.g. a standard ``box'' distribution),
then Gaussian tails will emerge due to the fluctuations in $\delta\Delta_{c}^{R}$,
and the ``size'' of the tails will be determined by an effective
disorder corresponding to $W_{eff}^{(0)}=W\sqrt{C}$. Here, the superscript $^{(0)}$
indicates that we ignored the interaction renormalizations. In the Appendix, we argue
that the effective scattering potentials $\Phi_{j}$
will further renormalize the disorder distribution, but the Gaussian tails will remain
as its generic feature.

\begin{figure}
\begin{center}\includegraphics[%
  width=2.919in,
  keepaspectratio]{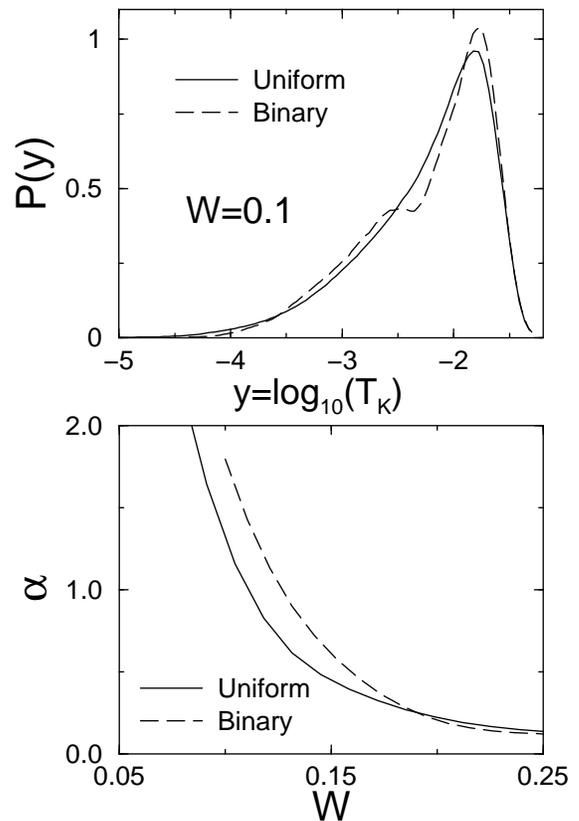}\end{center}

\caption{\label{fig5}\emph{Stat}DMFT results for the distribution of
Kondo temperatures for uniform and binary bare disorder distributions
(upper panel). The lower panel shows the exponent $\alpha$ as a function of disorder
strength. The results are qualitatively the same regardless of the specific form
of the disorder distribution. We used $V=0.5$, $E_{f}=-1$, and $\mu=-0.5$. }
\end{figure}

Now we present numerical results which provide strong evidence for the universality
of the renormalized disorder. Fig.~\ref{fig4} shows the results obtained within
the \emph{stat}DMFT\cite{mirandavlad1} for uniform and binary disorder distributions
with the same standard deviation $W=0.1$. As anticipated by Eq.~(\ref{4.8}), the fluctuations
of the cavity bath acquire an approximately Gaussian form with the same variance
for both bare disorder distributions, panel (b). The renormalized disorder distribution
$P(\widetilde{\varepsilon})$ exhibits long tails, panel (c), although the bare distributions
are bounded, panel (a). These Gaussian-like tails are the main universal feature
of the renormalized disorder, and they are crucial for the appearance of the singular
behavior in $P(T_{K})$ which leads to the formation of the Griffiths phase.

\emph{Stat}DMFT results in Fig.~\ref{fig5} provide further illustration of the
universality. The upper panel shows that the distributions of Kondo temperatures
for uniform and binary bare disorder distributions with the same standard deviation $W$
are qualitatively the same. The exponent $\alpha$, which determines the slope of
the distribution tails, is shown at the lower panel as a function of
$W$. It depends very weakly on the particular form of disorder distribution.

\subsection{Localization effects}

In the strict DMFT formulation $\alpha\approx t^{2}\rho_{o}J/2W^{2}$, where $\rho_{o}\equiv\rho_{av}$
is simply the (algebraic) average DOS of the conduction electrons, which therefore
remains finite even at the localization transition.\cite{andersonloc} If $J$ is
chosen to be large enough, the above seems to suggest that the Griffiths phase may
not emerge before the electrons localize at $W=W_{c}\sim1/\rho_{av}$. However, in
a theory that includes localization, the Kondo spins do not see the average, but
rather the \emph{typical} DOS of the conduction electrons.\cite{vladgabisdmft1,tmt}
Thus, in the NFL criterion, Eq.~(\ref{3.10}), one should actually replace $\rho_{o}\rightarrow\rho_{typ}$,
a quantity that becomes very small (and eventually vanishes) as the Anderson transition
is approached, viz. \begin{equation}
\rho_{typ}=A(W_{c}-W)^{\beta},\end{equation}
 where $A$ and $\beta$ are constants. We thus get \begin{equation}
W_{nfl}^{2}=\frac{1}{2}At^{2}J(W_{c}-W_{nfl})^{\beta}.\end{equation}
 This transcendental equation cannot be solved in closed form in general, but an
approximate solution can be found for $W_{c}^{2-\beta}/At^{2}J\ll1$. In this case,
the quantity $\delta W=1-W_{nfl}/W_{c}$ will be small, and to leading
order in $\delta W$\begin{equation}
W_{nfl}=W_{c}-(At^{2}J/2)^{-1/\beta}W_{c}^{2/\beta}<W_{c}.\end{equation}
 Therefore, the Griffiths phase emerges strictly \emph{before} the transition
is reached.

\section{Electronic Griffiths phase in the vicinity of the metal-insulator transition}

\label{sec:Electronic-Griffiths-phase}

Previous work\cite{vladgabisdmft1} has shown that the electronic
Griffiths phase appears also in a single band Hubbard model, as a
precursor to the Mott-Anderson metal-insulator transition (MIT).
Since the Hubbard model at half-filling is equivalent to the
charge-transfer model\cite{zaanenetal} of the MIT, we examine in
this Section the appearance of the Griffiths phase within this
model, which can be considered a version of the Anderson lattice
model that we examined in our approach.

\begin{figure}
\begin{center}\includegraphics[%
  width=2.919in,
  keepaspectratio]{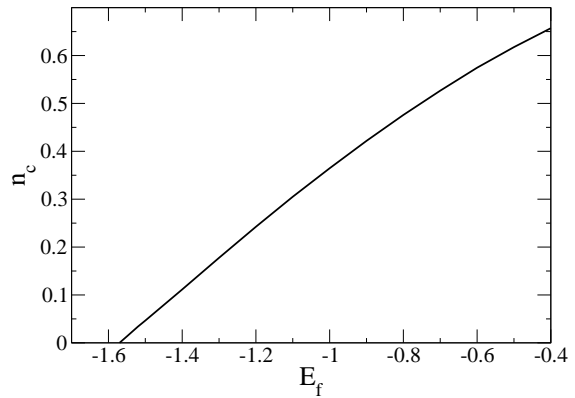}\end{center}

\caption{\label{fig6}Number of conduction electrons per site as a function of the $f$-level
energy $E_{f}$. The approach to the insulating phase is linear, $n_{c}\sim E_{f}-E_{f}^{c}$,
where $E_{f}^{c}$ is the critical value for the $f$-level energy. The hybridization
$V=0.5$, and $E_{f}$ is measured with respect to the middle of the conduction band. }
\end{figure}

The charge transfer model has been used to describe the Mott
metal-insulator transition for various systems, including copper
oxides.\cite{zhangrice} It consists of a two band model, where one
of the bands is narrow, and has large on-site interaction $U$
(Copper $d$-band), while the other band is broad enough that
electron-electron interactions can be neglected (Oxygen $p$-band).
A disordered version of this model is also appropriate to describe
the Mott-Anderson transition\cite{vladgabisdmft1} in doped
semiconductors such as Si:P. Here, the narrow band
corresponds\cite{shklovskiiefros} to the impurity band of the
Phosphorus donors, while the broad one is the conduction band of
Silicon. This model is given exactly by the Hamiltonian of
Eq.~(\ref{2.1}), but supplemented by the constraint
\begin{equation}
\overline{n_{fj}}+\overline{n_{cj}}=1,\label{6.1}\end{equation}
 which can be enforced by adjusting the value of the chemical potential. Here $n_{fj}$ and $n_{cj}$
are the average number of conduction and $f$-electrons on site $j$, and the overbar
denotes the average over disorder. In the mean field slave boson approach, the average
occupancy of the $f$-site is equal to \begin{equation}
n_{fj}=1-Z_{j}.\label{6.2}\end{equation}
 As the $f$-electron energy level is decreased ($|E_{f}|$ increased), the occupancy
of the $f$-sites becomes larger: the charge is {}``transferred''
from the conduction band. The transition to the Mott insulator is
found for sufficiently large $|E_{f}|$. At least within DMFT, this
metal-insulator transition has the same character as the more
familiar Mott transition in a single band Hubbard model. As an
illustration, we show on Fig.~\ref{fig6} the number of conduction
electrons per site, $n_{c}=Z$, as a function of $E_{f}$, in the
clean limit.

\begin{figure}
\begin{center}\includegraphics[%
  width=2.919in,
  keepaspectratio]{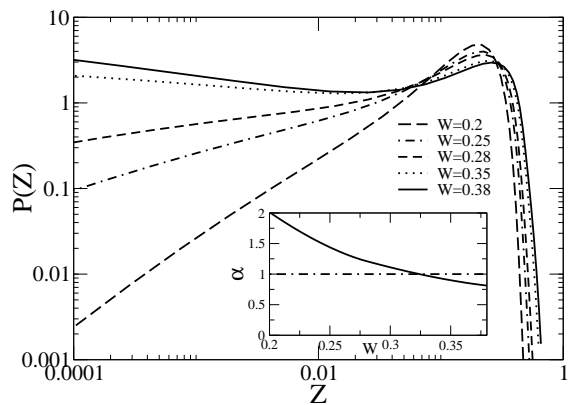}\end{center}

\caption{\label{fig7}Distribution of local Kondo temperatures for several levels of disorder.
The total number of electrons per unit cell is fixed to 1. The disorder ranges from
$W=0.2$ to $0.38$. We used $V=0.5$ and $E_{f}=-1.3$, where $E_{f}$ is measured
with respect to the middle of the conduction band. The inset shows the exponent $\alpha$
as a function of the disorder strength. The NFL phase occurs for $\alpha\leq1$. }
\end{figure}

We have solved our effective model in the parameter regime relevant to
the approach to the Mott-Anderson transition, and the results
demonstrate the emergence of an electronic Griffiths phase in the same
fashion as for the Anderson lattice model, consistent with \emph{
stat}DMFT results.\cite{vladgabisdmft1}. Here, the $f$-level energy in
Fig.~\ref{fig7} is measured with respect to the middle of the
conduction band, and not with respect to the chemical potential as in
Sec.~\ref{sec:Numerical-results}. For the parameters used in
Fig.~\ref{fig7}, the system is in the mixed valence regime, not in the
Kondo limit, and stronger disorder is needed for the appearance of the
NFL electronic Griffiths phase, again in close agreement with \emph{
stat}DMFT results.\cite{vladgabisdmft1} These results demonstrate that
our effective model proves capable to describe the emergence of the
electronic Griffiths phase as a universal phenomenon in correlated
electronic systems with disorder.

\section{Summary and outlook}

\label{sec:Summary-and-outlook}

In this paper we have identified an analytically solvable infinite
range model,
which captures the emergence of electronic Griffiths phases found
within the more generic \emph{stat}DMFT
approaches.\cite{mirandavlad1,aguiaretal1,vladgabisdmft1} In this
effective model, a specific distribution of disorder is
postulated, leading to a power-law distribution of local Kondo
temperatures and NFL behavior for sufficiently strong disorder. We
have also presented arguments explaining how this specific form of
randomness is universally generated by renormalizations due to
disorder-induced fluctuations of the conduction bath. In this
sense our effective model can be regarded as a (stable) fixed
point theory of electronic Griffiths phases.

The main motivation for introducing this effective model lies in
its simplicity, allowing an analytical solution, and thus
providing further insight into the mechanism for the emergence of
the electronic Griffiths phase. Nevertheless, an essential
ingredient is still missing from our Griffiths phase theory,
namely the inter-site RKKY interactions between Kondo spins.
According to the existing picture, all the spins with $T_K < T$
will not be Kondo screened, thus providing a large contribution to
thermodynamic response. These decoupled spins will, however, not
act as free local moments, but will feature dynamics dominated by
frustrating inter-site RKKY interactions. Recent experiments on
disordered Kondo alloys indeed seem to suggest the presence of
low temperature glassy dynamics with a negligible freezing
temperature.\cite{dougetal,dougetalreview} The simplifications
introduced by our effective model open an attractive avenue to
incorporate both the Kondo effect and the RKKY interaction in a
single theory. This fascinating direction remains a challenge for
future work.

\begin{acknowledgments}
We acknowledge fruitful discussions with Piers Coleman, Antoine
Georges, Qimiao Si, and Subir Sachdev. This work was supported by
FAPESP through grant 01/00719-8 (EM), by CNPq through grant
301222/97-5 (EM), and by the NSF  grant NSF-0234215 (VD and DT).
\end{acknowledgments}
\appendix



\section{Fluctuations in $\delta\Delta_{cj}^{I}$}

In Sec.~V we have ignored the fluctuations in the imaginary part of the cavity function
$\delta\Delta_{cj}^{I}.$ The corresponding contribution to the low-$T_{K}$ tail
is sub-leading, as we now show. We need to focus on rare events that produce exceptionally
small values of the local conduction electron DOS $\rho_{j}=-\frac{1}{\pi} G_{cj}(0)$.
Using Eq. (\ref{4.1}),
and ignoring the fluctuations in $\delta\Delta_{cj}^{R}$, we see that low values
for $\rho_{j}$ correspond to exceptionally high values for $\delta\Delta_{cj}^{I}.$
We therefore need to compute the form of the high-$\delta\Delta_{cj}^{I}$ tail of
$P(\delta\Delta_{cj}^{I}).$ Just as for the real part, we can estimate the fluctuations
of $\delta\Delta_{cj}^{I}$ by calculating the second moment,\begin{equation}
\sigma_{I}^{2}=\left\langle (\delta\Delta_{cj}^{I})^{2}\right\rangle ,\end{equation}
 and we get \begin{equation}
\sigma_{I}^{2}=C_{I}W^{2}+O(W^{4}),\end{equation}
 where\begin{equation}
C_{I}=\sum\limits _{\genfrac{}{}{0pt}{2}{j}{(j\neq0)}}
\left[ \operatorname{Im}\frac{(G_{oj})^{2}}{(G_{oo})^{2}}\right]^{2}.\end{equation}
 In this approximation, the quantity $\delta\Delta_{cj}^{I}$ has a Gaussian distribution,
and we find\begin{equation}
P(T_{K})\sim T_{K}^{-1}\exp\left\{ -\frac{J^{2}}{2\pi^{2}\sigma_{I}^{2}}\ln^{2}(D/T_{K})\right\} .\end{equation}
 As we can see, because  the ``log'' in the exponent has an extra
power of two, this distribution is log-normal and not power-law. Therefore the
$\delta\Delta_{cj}^{I}$
fluctuations, at least when treated on the Gaussian level as we have done, do not
lead to a singular $P(T_{K})$ distribution. Thus, to leading order we can ignore
these fluctuations when examining the emergence of the electronic
Griffiths phase.

\section{Interaction renormalizations}

In these estimates, we have omitted an important ingredient, the fact that Kondo
disorder itself will produce additional scattering, i.e. disorder in the conduction
channel, which needs to be self-consistently determined. As we have shown in previous
work,\cite{mirandavlad1} this results in a distribution of effective scattering
potentials $\Phi_{j}$, corresponding to the Kondo spins (note that in the uniform
case, the $\Phi_{j}$-s are the same on all sites, resulting in no scattering, but
contributing to the formation of the Kondo gap). The resulting scattering, in the
weak disorder limit again can be considered as a Gaussian distributed potential of
width\begin{equation}
W_{\Phi}=\left\langle \Phi_{j}^{2}\right\rangle ^{1/2}.\end{equation}
 Note however that this additional ``Kondo'' scattering does not
enter directly (at site $0$) in the solution of the local Kondo problem, since
the local f-site ``sees'' the corresponding c-site \textit{with the
f-site removed}. However, the presence of $\Phi_{j}$-s on all other sites $(j\neq0)$
does modify the form of $\delta\Delta_{cj}^{R}$ which, therefore, has to be computed
by including this additional scattering. At weak disorder, we expect\begin{equation}
\left\langle \Phi_{i}^{2}\right\rangle =C_{1}W_{eff}^{2},\end{equation}
where the constant $C_{1}$ measures the response of the Kondo spins to the hybridization
disorder. Note that $W_{eff}$ enters here, since the $\Phi_{j}$-s are obtained
from the solution of local Kondo problems, which are determined by the strength of
the renormalized site disorder, as modified by hybridization fluctuations. We therefore
need to compute $W_{eff}$ self-consistently, and we get\begin{equation}
W_{eff}^{2}=W^{2}+C(W^{2}+C_{2}W_{eff}^{2}),\end{equation}
 or\begin{equation}
W_{eff}=\sqrt{\frac{1+C}{1-CC_{2}}}W.\end{equation}

This reasoning, valid for weak bare disorder illustrates how the effective disorder
is generated in the conduction band even if it originally was not there, or is enhanced
due to additional Kondo scattering, if already present. In addition, these arguments
illustrate how Gaussian tails are generated for the renormalized disorder, even if
they are not introduced in the bare model. Of course, nonlinear effects at stronger
disorder cannot be accounted for in this simple fashion, which is especially true for
the consideration of the additional scattering introduced by disordered Kondo spins.
Nevertheless, the simple arguments that we presented illustrate how universality
is produced by renormalizations due to cavity field fluctuations, and seem to capture
the essential features of the emergence of the electronic Griffiths phase.


\begin{thebibliography}{34}
\expandafter\ifx\csname natexlab\endcsname\relax\def\natexlab#1{#1}\fi
\expandafter\ifx\csname bibnamefont\endcsname\relax
  \def\bibnamefont#1{#1}\fi
\expandafter\ifx\csname bibfnamefont\endcsname\relax
  \def\bibfnamefont#1{#1}\fi
\expandafter\ifx\csname citenamefont\endcsname\relax
  \def\citenamefont#1{#1}\fi
\expandafter\ifx\csname url\endcsname\relax
  \def\url#1{\texttt{#1}}\fi
\expandafter\ifx\csname urlprefix\endcsname\relax\def\urlprefix{URL }\fi
\providecommand{\bibinfo}[2]{#2}
\providecommand{\eprint}[2][]{\url{#2}}

\bibitem[{\citenamefont{{G. R. Stewart}}(2001)}]{stewartNFL}
\bibinfo{author}{\bibnamefont{{G. R. Stewart}}}, \bibinfo{journal}{Rev. Mod.
  Phys.} \textbf{\bibinfo{volume}{73}}, \bibinfo{pages}{797}
  (\bibinfo{year}{2001}).

\bibitem[{\citenamefont{{H. v. L\"{o}hneysen}}(1996)}]{lohneysen}
\bibinfo{author}{\bibnamefont{{H. v. L\"{o}hneysen}}}, \bibinfo{journal}{J.
  Phys.: Condens. Matter} \textbf{\bibinfo{volume}{8}}, \bibinfo{pages}{9689}
  (\bibinfo{year}{1996}).

\bibitem[{\citenamefont{{F. M. Grosche} et~al.}(2000)\citenamefont{{F. M.
  Grosche}, {P. Agarwal}, {S. R. Julian}, {N. J. Wilson}, {R. K. W.
  Haselwimmer}, {S. J. S. Lister}, {N. D. Mathur}, {F. V. Carter}, {S. S.
  Saxena}, and {G. G. Lonzarich}}}]{groscheetal}
\bibinfo{author}{\bibnamefont{{F. M. Grosche}}},
  \bibinfo{author}{\bibnamefont{{P. Agarwal}}},
  \bibinfo{author}{\bibnamefont{{S. R. Julian}}},
  \bibinfo{author}{\bibnamefont{{N. J. Wilson}}},
  \bibinfo{author}{\bibnamefont{{R. K. W. Haselwimmer}}},
  \bibinfo{author}{\bibnamefont{{S. J. S. Lister}}},
  \bibinfo{author}{\bibnamefont{{N. D. Mathur}}},
  \bibinfo{author}{\bibnamefont{{F. V. Carter}}},
  \bibinfo{author}{\bibnamefont{{S. S. Saxena}}}, \bibnamefont{and}
  \bibinfo{author}{\bibnamefont{{G. G. Lonzarich}}}, \bibinfo{journal}{J.
  Phys.: Condens. Matter} \textbf{\bibinfo{volume}{12}}, \bibinfo{pages}{L533}
  (\bibinfo{year}{2000}).

\bibitem[{\citenamefont{{J. A. Hertz}}(1976)}]{hertz}
\bibinfo{author}{\bibnamefont{{J. A. Hertz}}}, \bibinfo{journal}{Phys. Rev. B}
  \textbf{\bibinfo{volume}{14}}, \bibinfo{pages}{1165} (\bibinfo{year}{1976}).

\bibitem[{\citenamefont{{T. Moryia}}(1985)}]{moryia}
\bibinfo{author}{\bibnamefont{{T. Moryia}}}, \emph{\bibinfo{title}{Spin
  {F}luctuations in {I}tinerant {E}lectron {M}agnetism}}
  (\bibinfo{publisher}{Springer-Verlag}, \bibinfo{address}{Berlin},
  \bibinfo{year}{1985}).

\bibitem[{\citenamefont{{M. A. Continentino} et~al.}(1989)\citenamefont{{M. A.
  Continentino}, {G. M. Japiassu}, and {A. Troper}}}]{japiassuetal}
\bibinfo{author}{\bibnamefont{{M. A. Continentino}}},
  \bibinfo{author}{\bibnamefont{{G. M. Japiassu}}}, \bibnamefont{and}
  \bibinfo{author}{\bibnamefont{{A. Troper}}}, \bibinfo{journal}{Phys. Rev. B}
  \textbf{\bibinfo{volume}{39}}, \bibinfo{pages}{9734} (\bibinfo{year}{1989}).

\bibitem[{\citenamefont{{A. J. Millis}}(1993)}]{millis}
\bibinfo{author}{\bibnamefont{{A. J. Millis}}}, \bibinfo{journal}{Phys. Rev. B}
  \textbf{\bibinfo{volume}{48}}, \bibinfo{pages}{7183} (\bibinfo{year}{1993}).

\bibitem[{\citenamefont{{P. Coleman} et~al.}(2001)\citenamefont{{P. Coleman},
  {C. P\'epin}, {Q. Si}, and {R. Ramazashvili}}}]{pierspepinsirevaz}
\bibinfo{author}{\bibnamefont{{P. Coleman}}}, \bibinfo{author}{\bibnamefont{{C.
  P\'epin}}}, \bibinfo{author}{\bibnamefont{{Q. Si}}}, \bibnamefont{and}
  \bibinfo{author}{\bibnamefont{{R. Ramazashvili}}}, \bibinfo{journal}{J.
  Phys.: Condens. Matter} \textbf{\bibinfo{volume}{13}}, \bibinfo{pages}{R723}
  (\bibinfo{year}{2001}).

\bibitem[{\citenamefont{{A. Schr\"oder} et~al.}(2000)\citenamefont{{A.
  Schr\"oder}, {G. Aeppli}, {R. Coldea}, {M. Adams}, {O. Stockert}, {H.v.
  L\"ohneysen}, {E. Bucher}, {R. Ramazashvili}, and {P.
  Coleman}}}]{schroedernature}
\bibinfo{author}{\bibnamefont{{A. Schr\"oder}}},
  \bibinfo{author}{\bibnamefont{{G. Aeppli}}},
  \bibinfo{author}{\bibnamefont{{R. Coldea}}},
  \bibinfo{author}{\bibnamefont{{M. Adams}}}, \bibinfo{author}{\bibnamefont{{O.
  Stockert}}}, \bibinfo{author}{\bibnamefont{{H.v. L\"ohneysen}}},
  \bibinfo{author}{\bibnamefont{{E. Bucher}}},
  \bibinfo{author}{\bibnamefont{{R. Ramazashvili}}}, \bibnamefont{and}
  \bibinfo{author}{\bibnamefont{{P. Coleman}}}, \bibinfo{journal}{Nature}
  \textbf{\bibinfo{volume}{407}}, \bibinfo{pages}{351} (\bibinfo{year}{2000}).

\bibitem[{\citenamefont{{Q. Si} et~al.}(2001)\citenamefont{{Q. Si}, {S.
  Rabello}, {K. Ingersent}, and {J. L. Smith}}}]{sietal}
\bibinfo{author}{\bibnamefont{{Q. Si}}}, \bibinfo{author}{\bibnamefont{{S.
  Rabello}}}, \bibinfo{author}{\bibnamefont{{K. Ingersent}}}, \bibnamefont{and}
  \bibinfo{author}{\bibnamefont{{J. L. Smith}}}, \bibinfo{journal}{Nature}
  \textbf{\bibinfo{volume}{413}}, \bibinfo{pages}{804} (\bibinfo{year}{2001}).

\bibitem[{\citenamefont{{O. O. Bernal} et~al.}(1995)\citenamefont{{O. O.
  Bernal}, {D. E. MacLaughlin}, {H. G. Lukefahr}, and {B.
  Andraka}}}]{bernaletal}
\bibinfo{author}{\bibnamefont{{O. O. Bernal}}},
  \bibinfo{author}{\bibnamefont{{D. E. MacLaughlin}}},
  \bibinfo{author}{\bibnamefont{{H. G. Lukefahr}}}, \bibnamefont{and}
  \bibinfo{author}{\bibnamefont{{B. Andraka}}}, \bibinfo{journal}{Phys. Rev.
  Lett.} \textbf{\bibinfo{volume}{75}}, \bibinfo{pages}{2023}
  (\bibinfo{year}{1995}).

\bibitem[{\citenamefont{{C. H. Booth} et~al.}(1998)\citenamefont{{C. H. Booth},
  {D. E. MacLaughlin}, {R. H. Heffner}, {R. Chau}, {M. B. Maple}, and {G. H.
  Kwei}}}]{boothetal}
\bibinfo{author}{\bibnamefont{{C. H. Booth}}},
  \bibinfo{author}{\bibnamefont{{D. E. MacLaughlin}}},
  \bibinfo{author}{\bibnamefont{{R. H. Heffner}}},
  \bibinfo{author}{\bibnamefont{{R. Chau}}}, \bibinfo{author}{\bibnamefont{{M.
  B. Maple}}}, \bibnamefont{and} \bibinfo{author}{\bibnamefont{{G. H. Kwei}}},
  \bibinfo{journal}{Phys. Rev. Lett.} \textbf{\bibinfo{volume}{81}},
  \bibinfo{pages}{3960} (\bibinfo{year}{1998}).

\bibitem[{\citenamefont{{N. B\"uttgen} et~al.}(2000)\citenamefont{{N.
  B\"uttgen}, {W. Trinkl}, {J.-E. Weber}, {J. Hemberger}, {A. Loidl}, and {S.
  Kehrein}}}]{buttgenetal}
\bibinfo{author}{\bibnamefont{{N. B\"uttgen}}},
  \bibinfo{author}{\bibnamefont{{W. Trinkl}}},
  \bibinfo{author}{\bibnamefont{{J.-E. Weber}}},
  \bibinfo{author}{\bibnamefont{{J. Hemberger}}},
  \bibinfo{author}{\bibnamefont{{A. Loidl}}}, \bibnamefont{and}
  \bibinfo{author}{\bibnamefont{{S. Kehrein}}}, \bibinfo{journal}{Phys. Rev. B}
  \textbf{\bibinfo{volume}{62}}, \bibinfo{pages}{11545} (\bibinfo{year}{2000}).

\bibitem[{\citenamefont{{D. E. MacLaughlin} et~al.}(2001)\citenamefont{{D. E.
  MacLaughlin}, {O. O. Bernal}, {R. H. Heffner}, {G. J. Nieuwenhuys}, {M. S.
  Rose}, {J. E. Sonier}, {B. Andraka}, {R. Chau}, and {M. B.
  Maple}}}]{dougetal}
\bibinfo{author}{\bibnamefont{{D. E. MacLaughlin}}},
  \bibinfo{author}{\bibnamefont{{O. O. Bernal}}},
  \bibinfo{author}{\bibnamefont{{R. H. Heffner}}},
  \bibinfo{author}{\bibnamefont{{G. J. Nieuwenhuys}}},
  \bibinfo{author}{\bibnamefont{{M. S. Rose}}},
  \bibinfo{author}{\bibnamefont{{J. E. Sonier}}},
  \bibinfo{author}{\bibnamefont{{B. Andraka}}},
  \bibinfo{author}{\bibnamefont{{R. Chau}}}, \bibnamefont{and}
  \bibinfo{author}{\bibnamefont{{M. B. Maple}}}, \bibinfo{journal}{Phys. Rev.
  Lett.} \textbf{\bibinfo{volume}{87}}, \bibinfo{pages}{066402}
  (\bibinfo{year}{2001}).

\bibitem[{\citenamefont{{D. E. MacLaughlin} et~al.}(2002)\citenamefont{{D. E.
  MacLaughlin}, {O. O. Bernal}, {J. E. Sonier}, {R. H. Heffner}, {T.
  Taniguchi}, and {Y. Miyako}}}]{dougetal2}
\bibinfo{author}{\bibnamefont{{D. E. MacLaughlin}}},
  \bibinfo{author}{\bibnamefont{{O. O. Bernal}}},
  \bibinfo{author}{\bibnamefont{{J. E. Sonier}}},
  \bibinfo{author}{\bibnamefont{{R. H. Heffner}}},
  \bibinfo{author}{\bibnamefont{{T. Taniguchi}}}, \bibnamefont{and}
  \bibinfo{author}{\bibnamefont{{Y. Miyako}}}, \bibinfo{journal}{Phys. Rev. B}
  \textbf{\bibinfo{volume}{65}}, \bibinfo{pages}{184401}
  (\bibinfo{year}{2002}).

\bibitem[{\citenamefont{{E. D. Bauer} et~al.}(2002)\citenamefont{{E. D. Bauer},
  {C. H. Booth}, {G. H. Kwei}, {R. Chau}, and {M. B. Maple}}}]{baueretal}
\bibinfo{author}{\bibnamefont{{E. D. Bauer}}},
  \bibinfo{author}{\bibnamefont{{C. H. Booth}}},
  \bibinfo{author}{\bibnamefont{{G. H. Kwei}}},
  \bibinfo{author}{\bibnamefont{{R. Chau}}}, \bibnamefont{and}
  \bibinfo{author}{\bibnamefont{{M. B. Maple}}}, \bibinfo{journal}{Phys. Rev.
  B} \textbf{\bibinfo{volume}{65}}, \bibinfo{pages}{245114}
  (\bibinfo{year}{2002}).

\bibitem[{\citenamefont{{C. H. Booth} et~al.}(2002)\citenamefont{{C. H. Booth},
  {E.-W. Scheidt}, {U. Killer}, {A. Weber}, and {S. Kehrein}}}]{boothetal2}
\bibinfo{author}{\bibnamefont{{C. H. Booth}}},
  \bibinfo{author}{\bibnamefont{{E.-W. Scheidt}}},
  \bibinfo{author}{\bibnamefont{{U. Killer}}},
  \bibinfo{author}{\bibnamefont{{A. Weber}}}, \bibnamefont{and}
  \bibinfo{author}{\bibnamefont{{S. Kehrein}}}, \bibinfo{journal}{Phys. Rev. B}
  \textbf{\bibinfo{volume}{66}}, \bibinfo{pages}{140402}
  (\bibinfo{year}{2002}).

\bibitem[{\citenamefont{{E. Miranda} and {V.
  Dobrosavljevi\'{c}}}(2001{\natexlab{a}})}]{mirandavlad1}
\bibinfo{author}{\bibnamefont{{E. Miranda}}} \bibnamefont{and}
  \bibinfo{author}{\bibnamefont{{V. Dobrosavljevi\'{c}}}},
  \bibinfo{journal}{Phys. Rev. Lett.} \textbf{\bibinfo{volume}{86}},
  \bibinfo{pages}{264} (\bibinfo{year}{2001}{\natexlab{a}}).

\bibitem[{\citenamefont{{E. Miranda} and {V.
  Dobrosavljevi\'{c}}}(2001{\natexlab{b}})}]{mirandavlad2}
\bibinfo{author}{\bibnamefont{{E. Miranda}}} \bibnamefont{and}
  \bibinfo{author}{\bibnamefont{{V. Dobrosavljevi\'{c}}}}, \bibinfo{journal}{J.
  Magn. Magn. Mat.} \textbf{\bibinfo{volume}{226-230}}, \bibinfo{pages}{110}
  (\bibinfo{year}{2001}{\natexlab{b}}).

\bibitem[{\citenamefont{{M. C. O. Aguiar} et~al.}(2003)\citenamefont{{M. C. O.
  Aguiar}, {E. Miranda}, and {V. Dobrosavljevi\'{c}}}}]{aguiaretal1}
\bibinfo{author}{\bibnamefont{{M. C. O. Aguiar}}},
  \bibinfo{author}{\bibnamefont{{E. Miranda}}}, \bibnamefont{and}
  \bibinfo{author}{\bibnamefont{{V. Dobrosavljevi\'{c}}}},
  \bibinfo{journal}{Phys. Rev. B} \textbf{\bibinfo{volume}{68}},
  \bibinfo{pages}{125104} (\bibinfo{year}{2003}).

\bibitem[{\citenamefont{{V. Dobrosavljevi\'{c}} and {G.
  Kotliar}}(1997)}]{vladgabisdmft1}
\bibinfo{author}{\bibnamefont{{V. Dobrosavljevi\'{c}}}} \bibnamefont{and}
  \bibinfo{author}{\bibnamefont{{G. Kotliar}}}, \bibinfo{journal}{Phys. Rev.
  Lett.} \textbf{\bibinfo{volume}{78}}, \bibinfo{pages}{3943}
  (\bibinfo{year}{1997}).

\bibitem[{\citenamefont{{A. Georges} et~al.}(1996)\citenamefont{{A. Georges},
  {G. Kotliar}, {W. Krauth}, and {M. J. Rozenberg}}}]{georgesrmp}
\bibinfo{author}{\bibnamefont{{A. Georges}}}, \bibinfo{author}{\bibnamefont{{G.
  Kotliar}}}, \bibinfo{author}{\bibnamefont{{W. Krauth}}}, \bibnamefont{and}
  \bibinfo{author}{\bibnamefont{{M. J. Rozenberg}}}, \bibinfo{journal}{Rev.
  Mod. Phys.} \textbf{\bibinfo{volume}{68}}, \bibinfo{pages}{13}
  (\bibinfo{year}{1996}).

\bibitem[{\citenamefont{{D. E. MacLaughlin} et~al.}(2004)\citenamefont{{D. E.
  MacLaughlin}, {R. H. Heffner}, {O. O. Bernal}, {K. Ishida}, {J. E. Sonier},
  {G. J. Nieuwenhuys}, {M. B. Maple}, and {G. R. Stewart}}}]{dougetalreview}
\bibinfo{author}{\bibnamefont{{D. E. MacLaughlin}}},
  \bibinfo{author}{\bibnamefont{{R. H. Heffner}}},
  \bibinfo{author}{\bibnamefont{{O. O. Bernal}}},
  \bibinfo{author}{\bibnamefont{{K. Ishida}}},
  \bibinfo{author}{\bibnamefont{{J. E. Sonier}}},
  \bibinfo{author}{\bibnamefont{{G. J. Nieuwenhuys}}},
  \bibinfo{author}{\bibnamefont{{M. B. Maple}}}, \bibnamefont{and}
  \bibinfo{author}{\bibnamefont{{G. R. Stewart}}},
  \bibinfo{journal}{J. Phys.:Condens. Matter 16, S4479 (2004)}  
    (\bibinfo{year}{2004}).

\bibitem[{\citenamefont{{N. Read} and {D. M. Newns}}(1983)}]{readnewns2}
\bibinfo{author}{\bibnamefont{{N. Read}}} \bibnamefont{and}
  \bibinfo{author}{\bibnamefont{{D. M. Newns}}}, \bibinfo{journal}{J. Phys. C}
  \textbf{\bibinfo{volume}{16}}, \bibinfo{pages}{L1055} (\bibinfo{year}{1983}).

\bibitem[{\citenamefont{{P. Coleman}}(1987)}]{colemanlong}
\bibinfo{author}{\bibnamefont{{P. Coleman}}}, \bibinfo{journal}{Phys. Rev. B}
  \textbf{\bibinfo{volume}{35}}, \bibinfo{pages}{5072} (\bibinfo{year}{1987}).

\bibitem[{\citenamefont{{E. Miranda} et~al.}(1996)\citenamefont{{E. Miranda},
  {V. Dobrosavljevi\'{c}}, and {G. Kotliar}}}]{mirandavladgabi1}
\bibinfo{author}{\bibnamefont{{E. Miranda}}}, \bibinfo{author}{\bibnamefont{{V.
  Dobrosavljevi\'{c}}}}, \bibnamefont{and} \bibinfo{author}{\bibnamefont{{G.
  Kotliar}}}, \bibinfo{journal}{J. Phys.: Condens. Matter}
  \textbf{\bibinfo{volume}{8}}, \bibinfo{pages}{9871} (\bibinfo{year}{1996}).

\bibitem[{\citenamefont{{A. C. Hewson}}(1993)}]{hewson}
\bibinfo{author}{\bibnamefont{{A. C. Hewson}}}, \emph{\bibinfo{title}{The
  {K}ondo {P}roblem to {H}eavy {F}ermions}} (\bibinfo{publisher}{Cambrige
  University Press}, \bibinfo{address}{Cambridge}, \bibinfo{year}{1993}).

\bibitem[{\citenamefont{{C. M. Varma} et~al.}(1989)\citenamefont{{C. M. Varma},
  {P. B. Littlewood}, {S. Schmitt-Rink}, {E. Abrahams}, and {A. E.
  Ruckenstein}}}]{varmamfl}
\bibinfo{author}{\bibnamefont{{C. M. Varma}}},
  \bibinfo{author}{\bibnamefont{{P. B. Littlewood}}},
  \bibinfo{author}{\bibnamefont{{S. Schmitt-Rink}}},
  \bibinfo{author}{\bibnamefont{{E. Abrahams}}}, \bibnamefont{and}
  \bibinfo{author}{\bibnamefont{{A. E. Ruckenstein}}}, \bibinfo{journal}{Phys.
  Rev. Lett.} \textbf{\bibinfo{volume}{63}}, \bibinfo{pages}{1996}
  (\bibinfo{year}{1989}).

\bibitem[{\citenamefont{{E. Miranda} et~al.}(1997)\citenamefont{{E. Miranda},
  {V. Dobrosavljevi\'{c}}, and {G. Kotliar}}}]{mirandavladgabi2}
\bibinfo{author}{\bibnamefont{{E. Miranda}}}, \bibinfo{author}{\bibnamefont{{V.
  Dobrosavljevi\'{c}}}}, \bibnamefont{and} \bibinfo{author}{\bibnamefont{{G.
  Kotliar}}}, \bibinfo{journal}{Phys. Rev. Lett.}
  \textbf{\bibinfo{volume}{78}}, \bibinfo{pages}{290} (\bibinfo{year}{1997}).

\bibitem[{\citenamefont{{P. W. Anderson}}(1958)}]{andersonloc}
\bibinfo{author}{\bibnamefont{{P. W. Anderson}}}, \bibinfo{journal}{Phys. Rev.}
  \textbf{\bibinfo{volume}{109}}, \bibinfo{pages}{1492} (\bibinfo{year}{1958}).

\bibitem[{\citenamefont{Dobrosavljevi\'c
  et~al.}(2003)\citenamefont{Dobrosavljevi\'c, Pastor, and Nikoli\'c}}]{tmt}
\bibinfo{author}{\bibfnamefont{V.}~\bibnamefont{Dobrosavljevi\'c}},
  \bibinfo{author}{\bibfnamefont{A.~A.} \bibnamefont{Pastor}},
  \bibnamefont{and} \bibinfo{author}{\bibfnamefont{B.~K.}
  \bibnamefont{Nikoli\'c}}, \bibinfo{journal}{Europhys. Lett.}
  \textbf{\bibinfo{volume}{62}}, \bibinfo{pages}{76} (\bibinfo{year}{2003}).

\bibitem[{\citenamefont{{J. Zaanen} et~al.}(1985)\citenamefont{{J. Zaanen}, {G.
  A. Sawatzky}, and {J. W. Allen}}}]{zaanenetal}
\bibinfo{author}{\bibnamefont{{J. Zaanen}}}, \bibinfo{author}{\bibnamefont{{G.
  A. Sawatzky}}}, \bibnamefont{and} \bibinfo{author}{\bibnamefont{{J. W.
  Allen}}}, \bibinfo{journal}{Phys. Rev. Lett.} \textbf{\bibinfo{volume}{55}},
  \bibinfo{pages}{418} (\bibinfo{year}{1985}).

\bibitem[{\citenamefont{{F. C. Zhang} and {T. M. Rice}}(1988)}]{zhangrice}
\bibinfo{author}{\bibnamefont{{F. C. Zhang}}} \bibnamefont{and}
  \bibinfo{author}{\bibnamefont{{T. M. Rice}}}, \bibinfo{journal}{Phys. Rev. B}
  \textbf{\bibinfo{volume}{37}}, \bibinfo{pages}{3759} (\bibinfo{year}{1988}).

\bibitem[{\citenamefont{{B. I. Shklovskii} and {A. L.
  Efros}}(1984)}]{shklovskiiefros}
\bibinfo{author}{\bibnamefont{{B. I. Shklovskii}}} \bibnamefont{and}
  \bibinfo{author}{\bibnamefont{{A. L. Efros}}},
  \emph{\bibinfo{title}{Electronic {P}roperties of {D}oped {S}emiconductors}}
  (\bibinfo{publisher}{Springer-Verlag}, \bibinfo{address}{Berlin},
  \bibinfo{year}{1984}).

\end{thebibliography}
\end{document}